\documentclass{class_nature}

\usepackage[utf8]{inputenc}
\usepackage[T1]{fontenc}
\usepackage{textcomp}
\usepackage{gensymb}
\usepackage{amsmath} 
\usepackage{xcolor}
\usepackage{graphicx}
\usepackage{xfrac}
\usepackage{soul} 
\usepackage{siunitx}
\usepackage{lineno}

\newcommand{\HfO}{HfO$_2$}

\newcommand{\SiO}{SiO$_2$}

\makeatletter
\DeclareRobustCommand\bfseriesitshape{%
  \not@math@alphabet\itshapebfseries\relax
  \fontseries\bfdefault
  \fontshape\itdefault
  \selectfont
}
\makeatother
\DeclareTextFontCommand{\textbfit}{\bfseriesitshape}

\title{One nanometer HfO$_2$-based ferroelectric tunnel junctions on silicon}

\author{Suraj S. Cheema,$^{1\ast\dagger}$ Nirmaan Shanker,$^{1,2\ast}$ Cheng-Hsiang Hsu,$^{2\ast}$ Adhiraj Datar,$^{1,2}$ Jongho Bae,$^{2}$ Daewoong Kwon,$^{2,3}$ Sayeef Salahuddin$^{2,4\dagger}$}

\begin{document}

\maketitle

\begin{affiliations}
 \item Department of Materials Science and Engineering, University of California, Berkeley, CA, USA
 \item Department of Electrical Engineering and Computer Sciences, University of California, Berkeley, CA, USA
 \item Present address: Department of Electrical Engineering, Inha University Yonghyeon Campus, Incheon, 22212, South Korea.
 \item Materials Sciences Division, Lawrence Berkeley National Laboratory, Berkeley, CA, USA
 \newline \textbf{$^\ast$These authors contributed equally to this work}
 \newline \textbf{$^\dagger$Correspondence to: s.cheema@berkeley.edu (S.S.C); sayeef@berkeley.edu (S.S.)}
\end{affiliations}

\newpage


\begin{abstract}

In ferroelectric materials, spontaneous symmetry breaking leads to a switchable electric polarization, which offers significant promise for nonvolatile memories\cite{Namlab_TED_2020}. In particular, ferroelectric tunnel junctions (FTJs) have emerged as a new resistive switching memory which exploit polarization-dependent tunnel current across a thin ferroelectric barrier\cite{Tsymbal_Science_2006,FTJreview_NatCommun_2014,FTJreview_AdvMater_2019,BaTiO3_Nature_2009}. Here we demonstrate FTJs with CMOS-compatible Zr-doped {\HfO} (Zr:{\HfO}) ferroelectric barriers of just 1 nm thickness, grown by atomic layer deposition on silicon. These 1 nm Zr:{\HfO} tunnel junctions exhibit large polarization-driven electroresistance (19000 $\%$), the largest value reported for {\HfO}-based FTJs. 
In addition, due to just a 1 nm ferroelectric barrier, these junctions provide large tunnel current (> 1 A cm$^{-2}$) at low read voltage, orders of magnitude larger than reported thicker {\HfO}-based FTJs\cite{FTJreview_Elsevier_2019}. Therefore, our proof-of-principle demonstration provides an approach to simultaneously overcome three major drawbacks of prototypical FTJs: a Si-compatible ultrathin ferroelectric, large electroresistance, and large read current for high-speed operation\cite{Namlab_Nanotech_2019}.
%

\end{abstract}

\newpage


Ferroelectric materials are of great technological interest for low-power logic transistors\cite{Salahuddin_NanoLett_2008,Khan_APL_2011,Kwon_EDL_2019,Kwon_EDL_2020} and nonvolatile memories\cite{Namlab_TED_2020} due to collectively-ordered electrical dipoles whose polarization can be switched under an applied voltage\cite{Lines_Oxford_1977}. Most ferroelectric research has traditionally focused on perovskite-structure oxides\cite{Dawber_RevModPhys_2005}. Perovskites however, suffer from various chemical, thermal, lattice, and interfacial oxide incompatibilities with silicon and modern semiconductor processes\cite{Schlom_MRSBull_2008}. Since the discovery of ferroelectricity in {\HfO}-based thin films in 2011\cite{Namlab_APL_2011}, fluorite-structure binary oxides have attracted considerable interest as they are compatible with complementary metal-oxide-semiconductor (CMOS) processes\cite{Park_MRSCommun_2018}. Accordingly, {\HfO}-based ferroelectric memory has received significant attention in recent years\cite{Park_MRSCommun_2018,Namlab_TED_2020}, primarily focused on charge-based ferroelectric random access memory (FeRAM) and ferroelectric field effect transistors (FeFETs)\cite{Namlab_MRSBull_2018}. Meanwhile, resistive-switching materials -- which exhibit electrically-induced resistance changes in metal–dielectric–metal junctions -- have emerged as promising candidates for novel beyond-CMOS data-centric computing paradigms\cite{Review_NatElectron_2018,Review_NatRevMater_2020,Namlab_Nanotech_2019}. In this context, ferroelectric tunnel junctions (FTJs) present a promising energy-efficient resistive switching memory\cite{Review_NatRevMater_2020,Namlab_Nanotech_2019,FTJreview_NatElectron_2020} as FTJs exploit the ferroic polarization functionality of the insulating barrier\cite{Tsymbal_Science_2006}. Voltage-controlled polarization-dependent tunneling through the ferroelectric layer (tunnel electroresistance, TER) can yield much larger ON/OFF conductance ratios\cite{FTJreview_NatCommun_2014,FTJreview_AdvMater_2019} than, for example, current-controlled magnetic tunnel junctions\cite{Review_NatRevMater_2020}, another two-terminal tunneling resistive switching device.


A critical requirement for FTJs is to achieve a sufficiently high tunneling current ($J_{on}$) at the ON state to ensure that a scaled device can be read rapidly, while still exhibiting a large TER ((J$_{on}$-J$_{off}$)/J$_{off}$ x 100$\%$)\cite{Namlab_Nanotech_2019}. Considering the large band gap of {\HfO} ($\sim$6 eV), this means that the thickness of {\HfO} in the FTJ will need to be reduced to the ultrathin limit. Tunnel junctions implementing CMOS-compatible {\HfO}-based ferroelectric barriers have been recently demonstrated\cite{MFM_APL_2016,Toshiba_VLSI_2016,Namlab_ESSDERC_2018}, but even three nanometer Zr-doped {\HfO} (Zr:{\HfO}) barriers were found to be too thick to obtain nano-ampere level current in micron-sized capacitors\cite{EpiHZOftj_PhysRevAppl_2019}. Here, we demonstrate FTJs utilizing 1 nm Zr:{\HfO} as the ferroelectric barrier, grown by atomic layer deposition (ALD) directly on silicon, thereby scaling down the tunnel barrier thickness almost to the nanoscale limit. We recently demonstrated robust ferroelectricity in such 1 nm thick Zr:{\HfO} films\cite{Cheema_Nature_2020}. Additionally, we showed that, opposing conventional scaling trends observed in perovskite ferroelectrics, ferroelectricity is enhanced rather than suppressed with decreasing thickness in Zr:{\HfO}\cite{Cheema_Nature_2020}. Accordingly, FTJs employing these 1 nm Zr:{\HfO} tunnel junctions exhibit large polarization-driven TER (19000$\%$), the largest reported for {\HfO}-based FTJs (Fig. 4, Extended Data Fig. 1). In addition, these FTJs demonstrate large tunnel current ($J_{on}$ > 1 A cm$^{-2}$ at low read voltage), orders of magnitude larger than reported thicker {\HfO}-based FTJs\cite{Toshiba_VLSI_2016,Namlab_ESSDERC_2018} (Fig. 4). The demonstration of robust FTJ operation with the ferroelectric barrier scaled down almost to the physical limit and simultaneous occurrence of large TER and large tunnel current show potential for the eventual adoption of {\HfO}-based FTJs for ultra-scaled memory applications.


One nanometer films of Zr:{\HfO} are grown by 10 cycles of ALD (4:1 Hf:Zr cycle ratio) on highly-doped silicon wafers (10$^{19}$ cm$^{-3}$), buffered with approximately one nanometer chemically-grown {\SiO}, and capped with 50 nm W metal (Fig. 1a). For reference, 10 ALD cycles correspond to approximately \SI{10}{\angstrom} thickness, as confirmed by synchrotron x-ray reflectivity (XRR, Extended Data Fig. 2a) and extensively characterized in our previous work on ultrathin Zr:{\HfO} films\cite{Cheema_Nature_2020}. Post-deposition annealing at 500$\degree$C with W capping is required to stabilize the ferroelectric orthorhombic phase (Pca2$_1$) via confinement strain\cite{Cheema_Nature_2020} rather than the other nonpolar fluorite-structure polymorphs. Synchrotron x-ray characterization of these 1 nm Zr:{\HfO} films confirms the presence of the polar orthorhombic structure and highly-oriented films via in-plane grazing incidence diffraction (GID, Extended Data Fig. 2b) and two-dimensional reciprocal space maps (Extended Data Fig. 2c), respectively. 


To establish polarization switching in 1 nm Zr:{\HfO}-based capacitors (Fig. 1a), piezoresponse force microscopy (PFM) switching spectroscopy demonstrate the presence of 180$\degree$ phase hysteresis (Fig. 1b), consistent with ferroelectricity. More detailed advanced scanning probe characterization previously demonstrated ferroelectricity these 1 nm Zr:{\HfO} films\cite{Cheema_Nature_2020}. Pulsed current-voltage (\textit{I-V$_{write}$}) measurements (Fig. 1c) - applying the same waveform structure as PFM spectroscopy -- demonstrate demonstrate saturating, abrupt hysteretic behavior, with consistent coercive (switching) voltage as the PFM phase loops, again characteristic of polarization-driven switching\cite{Chanthbouala_NatNano_2012}. Notably, the presence of closed hysteresis likely precludes ionic-driven mechanisms, which have been reported to result in open \textit{I-V$_{write}$} loops in {\HfO}-based junctions\cite{EpiHZOftj_AdvElectronMater_2019}, potentially due to irreversible oxygen vacancy migration. The lack of a forming step at high voltage required to observe resistive hysteretic switching also renders ionic-driven filamentary mechanisms unlikely. To further eliminate potentially confounding contributions in the 1 nm Zr:{\HfO} FTJs, voltage polarity-dependent pulsed \textit{I-V} measurements (Fig. 1c) demonstrate a resistive hysteresis sense independent of the sweep direction, inconsistent with filament formation in electrochemical resistive switching, and consistent polarization-driven resistive switching (Methods). The observed counter-clockwise \textit{I-V$_{write}$} hysteresis sense can be explained by polarization-induced barrier height modulation in the metal-ferroelectric-insulator-semiconductor (MFIS) structure (Extended Data Fig. 3). Indeed, linear \textit{I-V$_{read}$} measurements fit well to a model considering direct tunneling through a polarization-dependent trapezoidal potential barrier (Methods, Extended Data Fig. 4). Device area-independence of $J_{on}$ (Extended Data Fig. 5) further eliminates filamentary-type mechanisms\cite{Toshiba_VLSI_2016} (Methods). Therefore, although the confounding, and often synergistic\cite{EpiHZOftj_AdvElectronMater_2019} influence of various electrochemical phenomena intertwined with polarization switching cannot be completely eliminated, multiple \textit{I-V} signatures in these 1 nm Zr:{\HfO} FTJs indicate minimal contributions from the relevant ionic-driven mechanisms commonly attributed to amorphous hafnia\cite{Namlab_Nanotech_2019}. Considering the flexibility of \textit{I-V} measurements to provide multiple independent methods of isolating polarization-driven switching from non-ferroelectric origins, these studies indicate that \textit{I-V} characterization of FTJs can also serve as a diagnostic of ultrathin ferroelectricity and metrology technique for advanced nanoelectronics\cite{Metrology_NatElectron_2018}. \textit{I-V} measurements in tunnel junctions overcome shortcomings faced by conventional ferroelectric polarization-voltage (\textit{P-V}), PFM, and synchrotron X-ray techniques by leveraging tunnel currents, disentangling competing hysteretic mechanisms, and demonstrating polarization switching, respectively (Methods).


Regarding device performance in the 1 nm Zr:{\HfO} FTJs, polarization-dependent \textit{I-V} measurements demonstrate ON/OFF conductance ratios approaching 200 from linear \textit{I-V$_{read}$} (Fig. 2a-c) and 100 from pulsed \textit{I-V$_{write}$} hysteresis (Fig. 2d), surpassing the previous high-mark around 50\cite{Tokyo_EDTM_2017} observed for {\HfO}-based FTJs (Fig. 4a, Extended Data Fig. 1). Achieving two orders of magnitude ON/OFF conductance ratio not only surpasses all {\HfO}-based FTJ literature, but also matches epitaxial perovskite-based FTJs grown by pulsed laser deposition (PLD) on silicon (Fig. 4b). Notably, $J_{on}$, $\sim$ 100 nA $\mu$m$^{-2}$ measured at low read voltage (300 mV) (Fig. 2), is orders of magnitude larger than observed in reported {\HfO}-based FTJs employing thicker ferroelectric barriers\cite{Toshiba_VLSI_2016,Namlab_ESSDERC_2018} (Fig. 4, Extended Data Fig. 1). The read voltage is well below the coercive voltage of 1 nm Zr:{\HfO}, enabling non-destructive readout. The low $J_{on}$ (< 10 nA $\mu$m$^{-2}$) reported for {\HfO}-based FTJs -- due to the lack of an ultrathin Zr:{\HfO} ferroelectric layer only recently demonstrated\cite{Cheema_Nature_2020} -- prevents practical application into highly-scaled crossbar memories due to insufficient read current\cite{Namlab_Nanotech_2019}. Here, the 1 nm Zr:{\HfO} FTJs can maintain an ON/OFF of 10 -- higher than most {\HfO}-based FTJs on silicon reported thus far (Extended Data Fig. 1) -- for up to 10$^3$ cycles (Fig. 3a). Endurance cycling operates slightly below the optimal switching voltage ($\pm$ 2.5 V) to prevent dielectric interlayer ({\SiO}) breakdown, a common failure mechanism in {\HfO}-based FeFETs\cite{Namlab_MRSBull_2018}. The large TER window in these 1 nm Zr:{\HfO} FTJs affords operation at a lower voltage to enhance endurance at the slight expense of TER. This proof-of-principle 1 nm Zr:{\HfO} FTJ should motivate further work to optimize this trade-off between TER and endurance, perhaps by employing a higher-$\kappa$ dielectric interlayer to improve the field-distribution through the ferroelectric. Elimination of the dielectric interlayer, e.g. depositing Zr:{\HfO} on Ge substrates without a native oxide layer, could enhance the endurance\cite{KAIST_Nanotech_2018}, but would likely diminish the TER without the asymmetry provided by the composite barrier\cite{Theory_IEDM_2018} (Extended Data Fig. 1). Retention measurements demonstrate the large ON/OFF window exceeding 100 can be maintained for at least 10$^4$ seconds (Fig. 3b), consistent with piezoresponse retention reported in these 1 nm Zr:{\HfO} films\cite{Cheema_Nature_2020}.


Beyond the ferroelectric barrier, there are various physical mechanisms within the metal-ferroelectric-insulator-semiconductor (MFIS) structure than could potentially explain the largest observed polarization-driven TER in {\HfO}-based FTJs to date. For one, MFS structures employing doped semiconductor electrodes enable barrier width modulation via the ferroelectric field effect at the dielectric-semiconductor interface. Accordingly, MFS structures have demonstrated enhanced performance in perovskite-based FTJs\cite{FTJreview_AdvMater_2019}, and even {\HfO}-based FTJs on Ge\cite{KAIST_Nanotech_2018} (Extended Data Fig. 1). However, barrier width modulation from the ferroelectric field-effect is not expected to be a major contributor to the observed behavior in this work (Methods). The presence of the dielectric {\SiO} interlayer and the high silicon electrode doping level (10$^{19}$ cm$^{-3}$) is expected to partially screen the Zr:{\HfO} polarization from the silicon interface and result in a negligible depletion region, respectively. The dual ferroelectric-dielectric barrier present in the heterostructure is another candidate mechanism\cite{Theory_APL_2009}; recent modeling of composite barrier {\HfO}-based FTJs predict depolarization fields from the dielectric interlayer to enhance the tunneling asymmetry, and therefore the TER\cite{Theory_IEDM_2019}. Another {\HfO}-based FTJ model\cite{Theory_IEDM_2018}, which combines both of the aforementioned layering effects present in our FTJ structure, found MFIS to be superior to MFIM structures due to the large asymmetry in dielectric screening between the top and bottom electrodes, which can reduce the OFF state current without diminishing the ON state current. Future {\HfO}-based experimental studies should optimize composite ferroelectric-dielectric bilayers\cite{Toshiba_VLSI_2016,Namlab_ESSDERC_2018,Namlab_JEDS_2019,FTJreview_Elsevier_2019} for FTJ performance while remaining in the ultrathin regime to maintain sufficient ON current.

Besides electrostatic effects related to the heterostructure stacking, the nature of the ultrathin ferroelectric layer itself can also contribute to the enhanced polarization-driven TER\cite{FTJreview_AdvMater_2019}. Perovskite-based FTJs have typically shown larger TER\cite{FTJreview_NatCommun_2014,FTJreview_AdvMater_2019} than reported {\HfO}-based FTJs (Extended Data Fig. 1), potentially due to the enhanced ferroelectric film quality and crystallinity in high-temperature PLD-grown perovskite ferroelectric films on lattice-matched substrates\cite{BaTiO3_Nature_2009} compared to low-temperature ALD-grown polycrystalline films on Si. In fact, ALD-grown Zr:{\HfO} on Ge demonstrate larger TER compared to ALD-grown Zr:{\HfO} on Si\cite{KAIST_Nanotech_2018}, attributed to the enhanced Zr:{\HfO} ferroelectric orthorhombic phase fraction and film quality on Ge\cite{KAIST_Nanotech_2018}. A similar effect can be expected in these ultrathin Zr:{\HfO} films considering they are highly-oriented (Extended Data Fig. 2), in contrast to thicker polycrystalline films\cite{Cheema_Nature_2020}. Furthermore, these 1 nm Zr:{\HfO} films display enhanced ferroelectricity\cite{Cheema_Nature_2020}; the amplified polar distortions (structural gauges of polarization) in 1 nm films can increase TER considering the barrier height modulation is proportional to polarization\cite{Theory_PRL_2005}. The common trend of diminished polarization as thickness is diminished -- often observed in perovskite ferroelectrics\cite{Ahn_Science_2004} -- does not hold for these 1 nm Zr:{\HfO} films which demonstrate "reverse" size effects\cite{Cheema_Nature_2020}. Therefore, the unique ferroelectric properties of ultrathin Zr:{\HfO}\cite{Cheema_Nature_2020} can help overcome the trade-off between high electroresistance -- typically diminishes with decreasing thickness\cite{BaTiO3_Nature_2009,FTJreview_AdvMater_2019} -- and tunnel current -- increases with decreasing thickness -- which have previously plagued FTJs\cite{FTJreview_AdvMater_2019}. Accordingly, the TER achieved in this work is two orders of magnitude larger than 1 nm perovskite-based FTJs on lattice-matched substrates\cite{BaTiO3_Nature_2009} and even matches thicker epitaxial perovskite-based ferroelectrics grown on silicon (Fig. 4b).


In summary, we have demonstrated FTJs on silicon in which the Zr:{\HfO} ferroelectric thickness has been scaled down to 1 nm, equivalent to just two fluorite-structure unit cells. These FTJs simultaneously achieve both large TER and ON current -- a combination that has eluded {\HfO}-based FTJs thus far (Fig. 4a, Extended Data Fig. 1). Furthermore, the data retention and endurance characteristics in these ultrathin {\HfO}-based FTJs are comparable to those obtained with much thicker {\HfO}-based ferroelectric layers. The ability to scale the ferroelectric thickness in an FTJ down to almost the physical thickness limit on silicon and maintain polarization-driven resistive switching offers great potential for high-density memory technology. 


\clearpage
\begin{figure}
\begin{centering}
\includegraphics{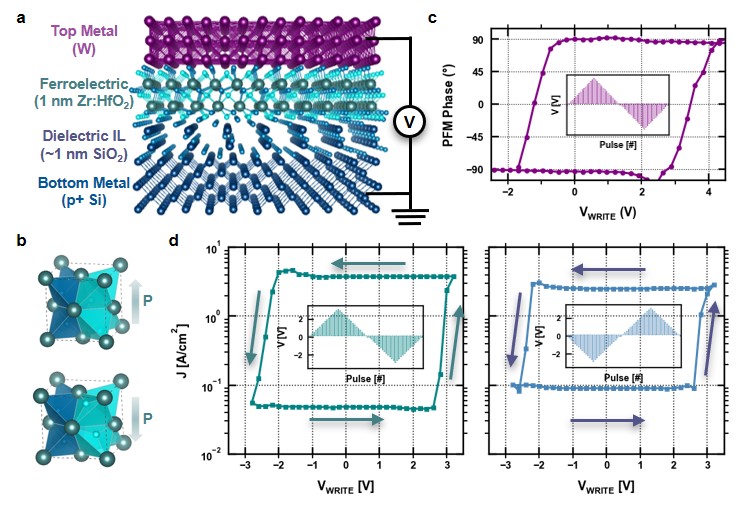}
\par\end{centering}
\end{figure}
\paragraph*{Fig. 1. Polarization-driven resistive switching in 1 nm Zr:{\HfO} junctions.}
{\bf (a)\/} Schematic of the Si/{\SiO}/Zr:{\HfO}/W FTJ heterostructure investigated in this work, utilizing composite tunnel barriers comprised of 1 nm crystalline ferroelectric Zr:{\HfO} and 1 nm amorphous dielectric {\SiO}. 
{\bf (b)\/} Crystal structure schematics of the polar orthorhombic fluorite-structure ferroelectric phase (Pca2$_1$), illustrating the two polarization states which dictate the resistive state in the FTJs. The different colored oxygen atoms represent the acentric oxygen atoms (cyan) and the centrosymmetric oxygen atoms (blue) within the surrounding cation (teal) tetrahedron.
{\bf (c)\/} Piezoresponse (phase) switching spectroscopy PFM loop for 1 nm Zr:{\HfO}, demonstrating ferroelectric-like hysteresis. More extensive scanning probe microscopy conclusively demonstrating ferroelectricity in these 1 nm Zr:{\HfO} films is reported in our previous work\cite{Cheema_Nature_2020}.
Inset: Voltage waveform used in PFM switching spectroscopy.
{\bf (d)\/} Pulsed current-voltage (\textit{I-V$_{write}$}) hysteresis map as a function of write voltage measured at 200 mV read voltage. The abrupt hysteretic behaviour and saturating tunnelling electroresistance is characteristic of polarization-driven switching, as opposed to filamentary-based switching caused by electrochemical migration (Methods). The device demonstrates voltage polarity-independent current–voltage hysteresis sense: both negative-positive-negative voltage polarity (left) and positive-negative-positive voltage polarity (right) demonstrate counter-clockwise hysteresis. The voltage polarity-dependent \textit{I-V} hysteresis measurements further rule out filamentary-based resistive switching mechanisms and is consistent with polarization-driven switching (Methods). 
Inset: Voltage waveform used in the pulsed \textit{I-V$_{write}$} measurements; the alternating sequence -- staircase write, fixed read -- mimics the PFM waveform. 
Band diagrams of the metal-ferroelectric-insulator-semiconductor (MFIS) heterostructure can explain the observed polarization-driven counterclockwise \textit{I-V$_{write}$} hysteresis (Extended Data Fig. 3), in which positive voltage applied to W results in a lower average potential barrier height, i.e. high-current state. 

\clearpage
\begin{figure}
\begin{centering}
\includegraphics{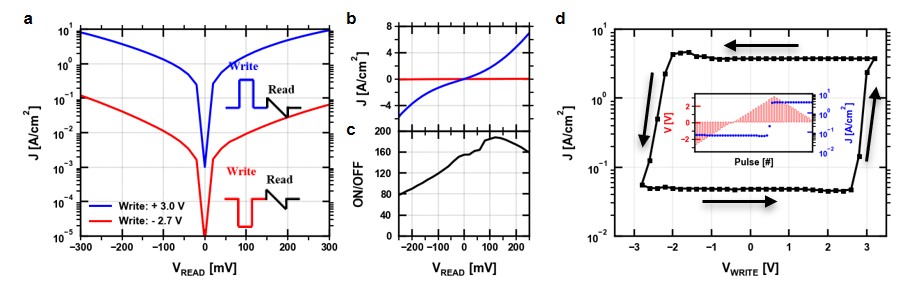}
\par\end{centering}
\end{figure}
\paragraph*{Fig. 2. Tunnel electroresistance in 1 nm Zr:{\HfO} ferroelectric tunnel junctions.}   
{\bf (a)\/} Linear \textit{I-V$_{read}$} measurements after the indicated write voltage (+3.0 V, -2.7 V) was applied to set the Zr:{\HfO} polarization (FTJ current) into its respective state. The \textit{I-V$_{read}$} data is consistent with a direct tunneling model considering polarization-dependent tunneling through a trapezoidal barrier (Methods, Extended Data Fig. 4).
Inset: Schematic of voltage waveform used to write and read the FTJ current states.
{\bf (b)\/} Tunnel current as a function of read voltage in linear scale to highlight the non-linear \textit{I-V$_{read}$} behavior.
{\bf (c)\/} Tunnel electroresistance (TER) ratio as a function of read voltage, demonstrating the maximum ON/OFF conductance ratio (190) is achieved around 150 mV.
{\bf (d)\/} Pulsed \textit{I-V$_{write}$} hysteresis map of the read current (measured at 200 mV) as a function of write voltage (up to $\pm$ 3 V), demonstrating saturating ON/OFF conductance ratio around 100.
Inset: Raw \textit{I-V} measurement as a function of pulse number to highlight the two stable current states in the FTJ corresponding to the two polarization states in the 1 nm Zr:{\HfO} ferroelectric layer.
The polarization-driven ON/OFF conductance ratio indicated from both steady-state \textbf{(a)} and pulsed \textbf{(d)} \textit{I-V} measurements far exceed the largest ratios reported thus far in {\HfO}-based FTJs on silicon (Fig. 4a) or any template (Extended Data Fig. 1), and even epitaxial perovskite-based FTJs grown on buffered silicon (Fig. 4b).

\clearpage
\begin{figure}
\begin{centering}
\includegraphics{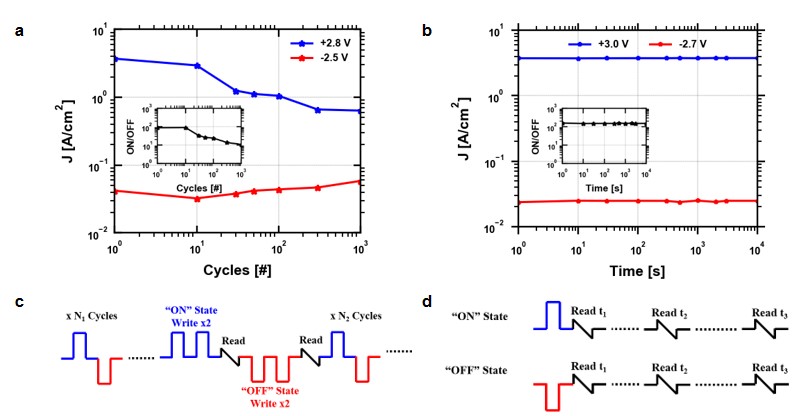}
\par\end{centering}
\end{figure}
\paragraph*{Fig. 3. Reliability of 1 nm Zr:{\HfO} ferroelectric tunnel junctions.}   
{\bf (a)\/} Read current (measured at 200 mV) as a function of endurance cycling (+2.8/-2.5 V write voltages) of the 1 nm Zr:{\HfO} FTJ.
Inset: ON/OFF ratio as a function of write cycles, which indicates 10x ON/OFF conductance ratio is maintained up to 10$^3$ cycles. The endurance of such a thin ferroelectric layer -- considering the extensive voltage drop through the amorphous dielectric interlayer (Methods) -- is enhanced by operating the FTJ at lower voltage, made possible by the large TER window.
{\bf (b)\/} Read current (measured at 200 mV) as a function of time after the FTJ is set into its respective state by the indicated write voltage (+3.0 V, -2.7 V). These retention measurements indicate the large TER can be maintained for up to 10$^4$ seconds, consistent with the lack of decay in our previously-reported 1 nm Zr:{\HfO} piezoresponse retention measurements\cite{Cheema_Nature_2020}.
{\bf (c), (d)\/} Voltage waveform sequence detailing read and write steps for the endurance \textbf{(c)} and retention \textbf{(d)} measurements.

\clearpage
\begin{figure}
\begin{centering}
\includegraphics{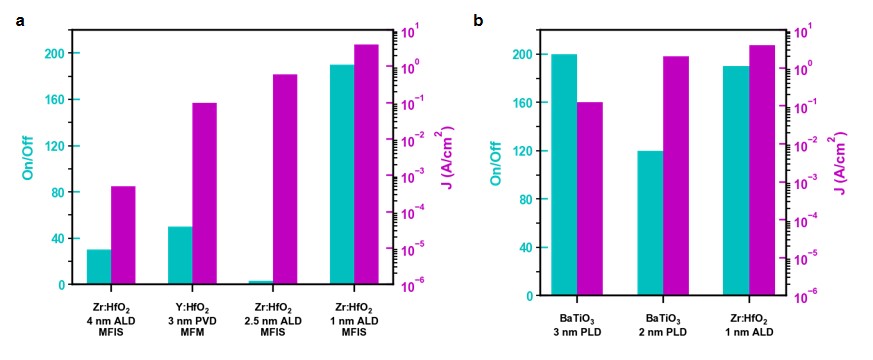}
\par\end{centering}
\end{figure}
\paragraph*{Fig. 4. Comparison to {\HfO}-based and perovskite-based FTJs on Si.}   
{\bf (a), (b)\/} Comparison of the 1 nm Zr:{\HfO} FTJs in this work to reported {\HfO}-based FTJs {\bf(a)} and perovskite-based FTJs {\bf(b)} grown on Si, considering polarization-driven ON/OFF conductance ratio ($J_{on}$/$J_{off}$) and current density ($J_{on}$). The ferroelectric barriers, thickness, deposition technique, and FTJ structure are indicated. The 1-nm Zr:{\HfO} FTJs presented in this work demonstrate the largest polarization-driven TER and current density reported thus far for {\HfO}-based FTJ literature on silicon {\bf(a)}, across all templates beyond just silicon (Extended Data Fig. 1), and even considering epitaxial perovskite ferroelectrics grown on silicon {\bf(b)}. Two orders of magnitude ON/OFF conductance ratio is the typical range observed for PLD-grown epitaxial perovskite-based FTJs\cite{FTJreview_AdvMater_2019}. $J_{on}$ is reported at 200 mV read voltage, except for the Y:{\HfO} (3 nm)\cite{Tokyo_EDTM_2017} case (500 mV). For this work, $J_{on}$ is consistent across various capacitor areas (Extended Data Fig. 5). The Y:{\HfO} (3 nm)\cite{Tokyo_EDTM_2017} example is deposited by sputtering (PVD), while the Zr:{\HfO} (2.5 nm)\cite{Moscow_MicroEng_2017}, Zr:{\HfO} (4 nm)\cite{Tokyo_JEDS_2019} and Zr:{\HfO} (1 nm, this work) ferroelectrics are deposited by the CMOS-compatible ALD technique. The BaTiO$_3$ (2 nm)\cite{FTJonSi_SciRep_2015} and BaTiO$_3$ (3 nm)\cite{FTJonSi_AdvMater_2014} examples are grown by PLD on thick perovskite-buffered Si/SiO$_x$ (SrTiO$_3$ perovskite transition layer, La$_{1-x}$Sr$_x$MnO$_3$ bottom electrode), which would prevent vertical-scaling, as well as introduce other CMOS-compatibility issues. Both high TER and high $J_{on}$ are required for highly-scaled FTJ application\cite{FTJreview_Elsevier_2019}, and ultrathin ferroelectric barriers enable emerging high-density three-dimensional computing architectures\cite{Review_NatRevMater_2020}. 


\newpage
\bibliographystyle{naturemag}
\bibliography{Refs_All}

\begin{thebibliography}{10}
\expandafter\ifx\csname url\endcsname\relax
  \def\url#1{\texttt{#1}}\fi
\expandafter\ifx\csname urlprefix\endcsname\relax\def\urlprefix{URL }\fi
\providecommand{\bibinfo}[2]{#2}
\providecommand{\eprint}[2][]{\url{#2}}

\bibitem{Namlab_TED_2020}
\bibinfo{author}{Mikolajick, T.}, \bibinfo{author}{Schroeder, U.} \&
  \bibinfo{author}{Slesazeck, S.}
\newblock \bibinfo{title}{{The Past, the Present, and the Future of
  Ferroelectric Memories}}.
\newblock \emph{\bibinfo{journal}{IEEE Transactions on Electron Devices}}
  \textbf{\bibinfo{volume}{67}}, \bibinfo{pages}{1434--1443}
  (\bibinfo{year}{2020}).

\bibitem{Tsymbal_Science_2006}
\bibinfo{author}{Tsymbal, E.~Y.} \& \bibinfo{author}{{H. Kohlstedt}}.
\newblock \bibinfo{title}{{Tunneling Across a Ferroelectric}}.
\newblock \emph{\bibinfo{journal}{Science}} \textbf{\bibinfo{volume}{313}},
  \bibinfo{pages}{181--183} (\bibinfo{year}{2006}).

\bibitem{FTJreview_NatCommun_2014}
\bibinfo{author}{Garcia, V.} \& \bibinfo{author}{Bibes, M.}
\newblock \bibinfo{title}{{Ferroelectric tunnel junctions for information
  storage and processing}}.
\newblock \emph{\bibinfo{journal}{Nature Communications}}
  \textbf{\bibinfo{volume}{5}}, \bibinfo{pages}{4289} (\bibinfo{year}{2014}).

\bibitem{FTJreview_AdvMater_2019}
\bibinfo{author}{Wen, Z.} \& \bibinfo{author}{Wu, D.}
\newblock \bibinfo{title}{{Ferroelectric Tunnel Junctions: Modulations on the
  Potential Barrier}}.
\newblock \emph{\bibinfo{journal}{Advanced Materials}} \bibinfo{pages}{1904123}
  (\bibinfo{year}{2019}).

\bibitem{BaTiO3_Nature_2009}
\bibinfo{author}{Garcia, V.} \emph{et~al.}
\newblock \bibinfo{title}{{Giant tunnel electroresistance for non-destructive
  readout of ferroelectric states}}.
\newblock \emph{\bibinfo{journal}{Nature}} \textbf{\bibinfo{volume}{460}},
  \bibinfo{pages}{81--84} (\bibinfo{year}{2009}).

\bibitem{FTJreview_Elsevier_2019}
\bibinfo{author}{Fujii, S.} \& \bibinfo{author}{Saitoh, M.}
\newblock \bibinfo{title}{{Ferroelectric Tunnel Junction}}.
\newblock In \emph{\bibinfo{booktitle}{Ferroelectricity in Doped Hafnium Oxide:
  Materials, Properties and Devices}}, \bibinfo{pages}{437--449}
  (\bibinfo{publisher}{Elsevier}, \bibinfo{year}{2019}).

\bibitem{Namlab_Nanotech_2019}
\bibinfo{author}{Slesazeck, S.} \& \bibinfo{author}{Mikolajick, T.}
\newblock \bibinfo{title}{{Nanoscale resistive switching memory devices: a
  review}}.
\newblock \emph{\bibinfo{journal}{Nanotechnology}}
  \textbf{\bibinfo{volume}{30}}, \bibinfo{pages}{352003}
  (\bibinfo{year}{2019}).

\bibitem{Salahuddin_NanoLett_2008}
\bibinfo{author}{Salahuddin, S.} \& \bibinfo{author}{Datta, S.}
\newblock \bibinfo{title}{{Use of Negative Capacitance to Provide Voltage
  Amplification for Low Power Nanoscale Devices}}.
\newblock \emph{\bibinfo{journal}{Nano Lett.}} \textbf{\bibinfo{volume}{8}},
  \bibinfo{pages}{405--410} (\bibinfo{year}{2008}).

\bibitem{Khan_APL_2011}
\bibinfo{author}{Khan, A.} \emph{et~al.}
\newblock \bibinfo{title}{{Experimental evidence of ferroelectric negative
  capacitance in nanoscale heterostructures}}.
\newblock \emph{\bibinfo{journal}{Appl. Phys. Lett.}}
  \textbf{\bibinfo{volume}{99}}, \bibinfo{pages}{113501}
  (\bibinfo{year}{2011}).

\bibitem{Kwon_EDL_2019}
\bibinfo{author}{Kwon, D.} \emph{et~al.}
\newblock \bibinfo{title}{{Negative Capacitance FET With 1.8-nm-Thick Zr-Doped
  HfO$_2$ Oxide}}.
\newblock \emph{\bibinfo{journal}{IEEE Electron Device Letters}}
  \textbf{\bibinfo{volume}{40}}, \bibinfo{pages}{993--996}
  (\bibinfo{year}{2019}).

\bibitem{Kwon_EDL_2020}
\bibinfo{author}{Kwon, D.} \emph{et~al.}
\newblock \bibinfo{title}{{Near Threshold Capacitance Matching in a Negative
  Capacitance FET With 1 nm Effective Oxide Thickness Gate Stack}}.
\newblock \emph{\bibinfo{journal}{IEEE Electron Device Letters}}
  \textbf{\bibinfo{volume}{41}}, \bibinfo{pages}{179--182}
  (\bibinfo{year}{2020}).

\bibitem{Lines_Oxford_1977}
\bibinfo{author}{Lines, M.~E.} \& \bibinfo{author}{Glass, A.~M.}
\newblock \emph{\bibinfo{title}{Principles and applications of ferroelectrics
  and related materials}} (\bibinfo{publisher}{Oxford University Press},
  \bibinfo{year}{1977}).

\bibitem{Dawber_RevModPhys_2005}
\bibinfo{author}{Dawber, M.}, \bibinfo{author}{Rabe, K.~M.} \&
  \bibinfo{author}{Scott, J.~F.}
\newblock \bibinfo{title}{Physics of thin-film ferroelectric oxides}.
\newblock \emph{\bibinfo{journal}{Rev. Mod. Phys.}}
  \textbf{\bibinfo{volume}{77}}, \bibinfo{pages}{1083--1130}
  (\bibinfo{year}{2005}).

\bibitem{Schlom_MRSBull_2008}
\bibinfo{author}{Schlom, D.~G.}, \bibinfo{author}{Guha, S.} \&
  \bibinfo{author}{Datta, S.}
\newblock \bibinfo{title}{{Gate Oxides Beyond SiO$_2$}}.
\newblock \emph{\bibinfo{journal}{{MRS} Bulletin}}
  \textbf{\bibinfo{volume}{33}}, \bibinfo{pages}{1017--1025}
  (\bibinfo{year}{2008}).

\bibitem{Namlab_APL_2011}
\bibinfo{author}{B{\"o}scke, T.~S.}, \bibinfo{author}{M{\"u}ller, J.},
  \bibinfo{author}{Br{\"a}uhaus, D.}, \bibinfo{author}{Schr{\"o}der, U.} \&
  \bibinfo{author}{B{\"o}ttger, U.}
\newblock \bibinfo{title}{Ferroelectricity in hafnium oxide thin films}.
\newblock \emph{\bibinfo{journal}{Appl. Phys. Lett.}}
  \textbf{\bibinfo{volume}{99}}, \bibinfo{pages}{102903}
  (\bibinfo{year}{2011}).

\bibitem{Park_MRSCommun_2018}
\bibinfo{author}{Park, M.}, \bibinfo{author}{Lee, Y.},
  \bibinfo{author}{Mikolajick, T.}, \bibinfo{author}{Schroeder, U.} \&
  \bibinfo{author}{Hwang, C.}
\newblock \bibinfo{title}{{Review and perspective on ferroelectric
  HfO$_2$-based thin films for memory applications}}.
\newblock \emph{\bibinfo{journal}{{MRS Commun.}}} \bibinfo{pages}{1--14}
  (\bibinfo{year}{2018}).

\bibitem{Namlab_MRSBull_2018}
\bibinfo{author}{Mikolajick, T.}, \bibinfo{author}{Slesazeck, S.},
  \bibinfo{author}{Park, M.} \& \bibinfo{author}{Schroeder, U.}
\newblock \bibinfo{title}{Ferroelectric hafnium oxide for ferroelectric
  random-access memories and ferroelectric field-effect transistors}.
\newblock \emph{\bibinfo{journal}{{MRS Bulletin}}}
  \textbf{\bibinfo{volume}{43}}, \bibinfo{pages}{340--346}
  (\bibinfo{year}{2018}).

\bibitem{Review_NatElectron_2018}
\bibinfo{author}{Ielmini, D.} \& \bibinfo{author}{Wong, H.-S.~P.}
\newblock \bibinfo{title}{{In-memory computing with resistive switching
  devices}}.
\newblock \emph{\bibinfo{journal}{Nature Electronics}}
  \textbf{\bibinfo{volume}{1}}, \bibinfo{pages}{333--343}
  (\bibinfo{year}{2018}).

\bibitem{Review_NatRevMater_2020}
\bibinfo{author}{Wang, Z.} \emph{et~al.}
\newblock \bibinfo{title}{{Resistive switching materials for information
  processing}}.
\newblock \emph{\bibinfo{journal}{Nature Reviews Materials}}
  \textbf{\bibinfo{volume}{5}}, \bibinfo{pages}{173--195}
  (\bibinfo{year}{2020}).

\bibitem{FTJreview_NatElectron_2020}
\bibinfo{author}{Yang, R.}
\newblock \bibinfo{title}{{In-memory computing with ferroelectrics}}.
\newblock \emph{\bibinfo{journal}{Nature Electronics}}  (\bibinfo{year}{2020}).

\bibitem{MFM_APL_2016}
\bibinfo{author}{Fan, Z.} \emph{et~al.}
\newblock \bibinfo{title}{{Ferroelectricity and ferroelectric resistive
  switching in sputtered Hf$_{0.5}$Zr$_{0.5}$O$_2$ thin films}}.
\newblock \emph{\bibinfo{journal}{Applied Physics Letters}}
  \textbf{\bibinfo{volume}{108}}, \bibinfo{pages}{232905}
  (\bibinfo{year}{2016}).

\bibitem{Toshiba_VLSI_2016}
\bibinfo{author}{Fujii, S.} \emph{et~al.}
\newblock \bibinfo{title}{{First demonstration and performance improvement of
  ferroelectric HfO$_2$-based resistive switch with low operation current and
  intrinsic diode property}}.
\newblock In \emph{\bibinfo{booktitle}{2016 IEEE Symposium on VLSI
  Technology}}, \bibinfo{pages}{1--2} (\bibinfo{publisher}{IEEE},
  \bibinfo{year}{2016}).

\bibitem{Namlab_ESSDERC_2018}
\bibinfo{author}{Max, B.}, \bibinfo{author}{Hoffmann, M.},
  \bibinfo{author}{Slesazeck, S.} \& \bibinfo{author}{Mikolajick, T.}
\newblock \bibinfo{title}{{Ferroelectric Tunnel Junctions based on
  Ferroelectric-Dielectric Hf$_{0.5}$Zr$_{0.5}$O$_2$/Al$_2$O$_3$ Capacitor
  Stacks}}.
\newblock \emph{\bibinfo{journal}{2018 48th European Solid-State Device
  Research Conference (ESSDERC)}} \bibinfo{pages}{142--145}
  (\bibinfo{year}{2018}).

\bibitem{EpiHZOftj_PhysRevAppl_2019}
\bibinfo{author}{Wei, Y.} \emph{et~al.}
\newblock \bibinfo{title}{{Magnetic Tunnel Junctions Based on Ferroelectric
  Hf$_{0.5}$Zr$_{0.5}$O$_2$ Tunnel Barriers}}.
\newblock \emph{\bibinfo{journal}{Physical Review Applied}}
  \textbf{\bibinfo{volume}{12}}, \bibinfo{pages}{031001}
  (\bibinfo{year}{2019}).

\bibitem{Cheema_Nature_2020}
\bibinfo{author}{Cheema, S.~S.} \emph{et~al.}
\newblock \bibinfo{title}{{Enhanced ferroelectricity in ultrathin films grown
  directly on silicon}}.
\newblock \emph{\bibinfo{journal}{Nature}} \textbf{\bibinfo{volume}{580}},
  \bibinfo{pages}{478--482} (\bibinfo{year}{2020}).

\bibitem{Chanthbouala_NatNano_2012}
\bibinfo{author}{Chanthbouala, A.} \emph{et~al.}
\newblock \bibinfo{title}{{Solid-state memories based on ferroelectric tunnel
  junctions}}.
\newblock \emph{\bibinfo{journal}{Nature Nanotechnology}}
  \textbf{\bibinfo{volume}{7}}, \bibinfo{pages}{101--104}
  (\bibinfo{year}{2012}).

\bibitem{EpiHZOftj_AdvElectronMater_2019}
\bibinfo{author}{Sulzbach, M.~C.} \emph{et~al.}
\newblock \bibinfo{title}{{Unraveling Ferroelectric Polarization and Ionic
  Contributions to Electroresistance in Epitaxial Hf$_{0.5}$Zr$_{0.5}$O$_2$
  Tunnel Junctions}}.
\newblock \emph{\bibinfo{journal}{Advanced Electronic Materials}}
  \bibinfo{pages}{1900852} (\bibinfo{year}{2019}).

\bibitem{Metrology_NatElectron_2018}
\bibinfo{author}{Orji, N.~G.} \emph{et~al.}
\newblock \bibinfo{title}{{Metrology for the next generation of semiconductor
  devices}}.
\newblock \emph{\bibinfo{journal}{Nature Electronics}}
  \textbf{\bibinfo{volume}{1}}, \bibinfo{pages}{532--547}
  (\bibinfo{year}{2018}).

\bibitem{Tokyo_EDTM_2017}
\bibinfo{author}{Tian, X.} \& \bibinfo{author}{Toriumi, A.}
\newblock \bibinfo{title}{{New opportunity of ferroelectric tunnel junction
  memory with ultrathin HfO$_2$-based oxides}}.
\newblock In \emph{\bibinfo{booktitle}{2017 IEEE Electron Devices Technology
  and Manufacturing Conference (EDTM)}}, \bibinfo{pages}{63--64}
  (\bibinfo{publisher}{IEEE}, \bibinfo{year}{2017}).

\bibitem{KAIST_Nanotech_2018}
\bibinfo{author}{Goh, Y.} \& \bibinfo{author}{Jeon, S.}
\newblock \bibinfo{title}{{The effect of the bottom electrode on ferroelectric
  tunnel junctions based on CMOS-compatible HfO$_2$}}.
\newblock \emph{\bibinfo{journal}{Nanotechnology}}
  \textbf{\bibinfo{volume}{29}}, \bibinfo{pages}{335201}
  (\bibinfo{year}{2018}).

\bibitem{Theory_IEDM_2018}
\bibinfo{author}{Mo, F.}, \bibinfo{author}{Tagawa, Y.},
  \bibinfo{author}{Saraya, T.}, \bibinfo{author}{Hiramoto, T.} \&
  \bibinfo{author}{Kobayashi, M.}
\newblock \bibinfo{title}{{Scalability Study on Ferroelectric-HfO2 Tunnel
  Junction Memory Based on Non-equilibrium Green Function Method with
  Self-consistent Potential}}.
\newblock In \emph{\bibinfo{booktitle}{2018 IEEE International Electron Devices
  Meeting (IEDM)}}, \bibinfo{pages}{16.3.1--16.3.4} (\bibinfo{publisher}{IEEE},
  \bibinfo{year}{2018}).

\bibitem{Theory_APL_2009}
\bibinfo{author}{Zhuravlev, M.~Y.}, \bibinfo{author}{Wang, Y.},
  \bibinfo{author}{Maekawa, S.} \& \bibinfo{author}{Tsymbal, E.~Y.}
\newblock \bibinfo{title}{{Tunneling electroresistance in ferroelectric tunnel
  junctions with a composite barrier}}.
\newblock \emph{\bibinfo{journal}{Applied Physics Letters}}
  \textbf{\bibinfo{volume}{95}}, \bibinfo{pages}{052902}
  (\bibinfo{year}{2009}).

\bibitem{Theory_IEDM_2019}
\bibinfo{author}{Huang, H.-H.} \emph{et~al.}
\newblock \bibinfo{title}{{A Comprehensive Modeling Framework for Ferroelectric
  Tunnel Junctions}}.
\newblock In \emph{\bibinfo{booktitle}{2019 IEEE International Electron Devices
  Meeting (IEDM)}}, \bibinfo{pages}{32.2.1--32.2.4} (\bibinfo{publisher}{IEEE},
  \bibinfo{year}{2019}).

\bibitem{Namlab_JEDS_2019}
\bibinfo{author}{Max, B.}, \bibinfo{author}{Hoffmann, M.},
  \bibinfo{author}{Slesazeck, S.} \& \bibinfo{author}{Mikolajick, T.}
\newblock \bibinfo{title}{{Direct Correlation of Ferroelectric Properties and
  Memory Characteristics in Ferroelectric Tunnel Junctions}}.
\newblock \emph{\bibinfo{journal}{IEEE Journal of the Electron Devices
  Society}} \textbf{\bibinfo{volume}{7}}, \bibinfo{pages}{1175--1181}
  (\bibinfo{year}{2019}).

\bibitem{Theory_PRL_2005}
\bibinfo{author}{Zhuravlev, M.~Y.}, \bibinfo{author}{Sabirianov, R.~F.},
  \bibinfo{author}{Jaswal, S.~S.} \& \bibinfo{author}{Tsymbal, E.~Y.}
\newblock \bibinfo{title}{{Giant Electroresistance in Ferroelectric Tunnel
  Junctions}}.
\newblock \emph{\bibinfo{journal}{Physical Review Letters}}
  \textbf{\bibinfo{volume}{94}}, \bibinfo{pages}{246802}
  (\bibinfo{year}{2005}).

\bibitem{Ahn_Science_2004}
\bibinfo{author}{Ahn, C.}, \bibinfo{author}{Rabe, K.} \&
  \bibinfo{author}{Triscone, J.}
\newblock \bibinfo{title}{{Ferroelectricity at the Nanoscale: Local
  Polarization in Oxide Thin Films and Heterostructures}}.
\newblock \emph{\bibinfo{journal}{Science}} \textbf{\bibinfo{volume}{303}},
  \bibinfo{pages}{488--491} (\bibinfo{year}{2004}).

\bibitem{Moscow_MicroEng_2017}
\bibinfo{author}{Chouprik, A.} \emph{et~al.}
\newblock \bibinfo{title}{{Electron transport across ultrathin ferroelectric
  Hf$_{0.5}$Zr$_{0.5}$O$_2$ films on Si}}.
\newblock \emph{\bibinfo{journal}{Microelectronic Engineering}}
  \textbf{\bibinfo{volume}{178}}, \bibinfo{pages}{250--253}
  (\bibinfo{year}{2017}).

\bibitem{Tokyo_JEDS_2019}
\bibinfo{author}{Kobayashi, M.}, \bibinfo{author}{Tagawa, Y.},
  \bibinfo{author}{Mo, F.}, \bibinfo{author}{Saraya, T.} \&
  \bibinfo{author}{Hiramoto, T.}
\newblock \bibinfo{title}{{Ferroelectric HfO$_2$ Tunnel Junction Memory With
  High TER and Multi-Level Operation Featuring Metal Replacement Process}}.
\newblock \emph{\bibinfo{journal}{IEEE Journal of the Electron Devices
  Society}} \textbf{\bibinfo{volume}{7}}, \bibinfo{pages}{134--139}
  (\bibinfo{year}{2019}).

\bibitem{FTJonSi_SciRep_2015}
\bibinfo{author}{Guo, R.} \emph{et~al.}
\newblock \bibinfo{title}{{Functional ferroelectric tunnel junctions on
  silicon}}.
\newblock \emph{\bibinfo{journal}{Scientific Reports}}
  \textbf{\bibinfo{volume}{5}}, \bibinfo{pages}{12576} (\bibinfo{year}{2015}).

\bibitem{FTJonSi_AdvMater_2014}
\bibinfo{author}{Li, Z.} \emph{et~al.}
\newblock \bibinfo{title}{{An Epitaxial Ferroelectric Tunnel Junction on
  Silicon}}.
\newblock \emph{\bibinfo{journal}{Advanced Materials}}
  \textbf{\bibinfo{volume}{26}}, \bibinfo{pages}{7185--7189}
  (\bibinfo{year}{2014}).

\end{thebibliography}


\newpage
\begin{addendum}

\item [Acknowledgements] 
This research was supported in part by the following: the Berkeley Center for Negative Capacitance Transistors (BCNCT); Applications and Systems-Driven Center for Energy-Efficient Integrated NanoTechnologies (ASCENT), one of the six centres in the Joint University Microelectronics Program (JUMP) initiative, a Semiconductor Research Corporation (SRC) program sponsored by Defense Advanced Research Projects Agency (DARPA); the DARPA Technologies for Mixed-mode Ultra Scaled Integrated Circuits (T-MUSIC) programme; the University of California Multicampus Research Programs and Initiatives (UC MRPI) project.
This research used resources of the Advanced Photon Source, a U.S. Department of Energy (DOE) Office of Science User Facility operated for the DOE Office of Science by Argonne National Laboratory under Contract No. DE-AC02-06CH11357. 
Use of the Stanford Synchrotron Radiation Light source, SLAC National Accelerator Laboratory, is supported by the U.S. Department of Energy, Office of Science, Office of Basic Energy Sciences under Contract No. DE-AC02-76SF00515.

\item [Author Contributions] 
S.S.C. and S.S. designed the research.
S.S.C. performed the ferroelectric film synthesis.
C.H.H. performed device fabrication.
N.S. performed scanning probe microscopy.
S.S.C., N.S., and C.H.H. performed synchrotron x-ray structural characterization. 
S.S.C., N.S. and A.D performed electrical measurements.
S.S.C. and S.S. co-wrote the manuscript.
S.S. supervised the research. 
All authors contributed to discussions and manuscript preparations. 

\item [Competing Interests] 
The authors declare that they have no competing financial interests.

\item [Correspondence] 
Correspondence and requests for materials should be addressed to \newline
S.S.C. (s.cheema@berkeley.edu) or S.S. (sayeef@berkeley.edu).

\end{addendum}


\end{document}